\documentclass[flushrt,preprint]{aastex61}       
\usepackage{graphicx}
\usepackage{url}
\usepackage{amsmath,amssymb}
\usepackage{physymb}
\usepackage{natbib}
\begin{document}
\sloppy

\title{Changing U.S. Extreme Temperature Statistics}
\shorttitle{U.S. Temperature Extremes}
\shortauthors{Finkel \& Katz}
\author{J.~M.~Finkel}
\affil{Department of Physics, Washington University, St. Louis, Mo. 63130}
\author{J.~I.~Katz}
\affil{Department of Physics and McDonnell Center for the Space Sciences,
Washington University, St. Louis, Mo. 63130}
\affil{Tel.: 314-935-6276; Facs: 314-935-6219}
\email{katz@wuphys.wustl.edu}
\begin{abstract}
The rise in global mean temperature is an incomplete description of warming.
For many purposes, including agriculture and human life, temperature
extremes may be more important than temperature means and changes in local
extremes may be more important than mean global changes.  We define a
nonparametric statistic to describe extreme temperature behavior by
quantifying the frequency of local daily all-time highs and lows, normalized
by their frequency in the null hypothesis of no climate change.  We average
this metric over 1218 weather stations in the 48 contiguous United States,
and find significantly fewer all-time lows than for the null hypothesis of
unchanging climate.  Record highs, by contrast, exhibit no significant
trend.  The metric is evaluated by Monte Carlo simulation for stationary and
warming temperature distributions, permitting comparison of the statistics
of historic temperature records with those of modeled behavior. 
\end{abstract}

\section{Introduction}
\label{sec:intro}
The steady accumulation of greenhouse gases, most importantly CO$_2$, in the
atmosphere and the warming trend it implies have been noted for more than a
century and are the subject of a large literature \citep{H06,HRSL10,IPCC5AR}.
The extensive observational data are complemented by numerical climate
simulations, in the form of General Circulation Models, whose unprecedented
detail and resolution have been enabled by the revolution in
high-performance computing.

On a day-to-day basis, we and our agriculture experience weather, not
climate.  The extremes of weather are important, independently of its mean
\citep{KB92,K93,N95,E97,F02,KT03,A06,P08,M09,AK10,RC11,L16}.
For example, the length of a growing season is often defined as the interval
between the last vernal and first autumnal frosts.  Heat stress on plants
and animals, including people, is determined by peak temperatures rather
than the mean temperature.  Extreme temperatures can change independently of
the mean temperature and its seasonal variation that we describe as climate.

All-time temperature highs and lows are particularly stressing.  We define
a nonparametric statistic that measures the frequency of all-time record
highs or lows at a site on a specified calendar day.  This statistic does
not depend on a choice of a threshold, necessarily arbitrary, for a
temperature excursion to be defined as extreme or stressing, but is an
intrinsic and nondimensional property of the statistics of daily temperature
variation.

In the null hypothesis of no climate change, the probability that the last
daily maximum (or minimum) in a series of $N$ such maxima (or minima) sets
an all-time record (assuming no ties) is $1/N$.  By comparing the actual
number of such all-time records in a database of 1218 sites in the 48
contiguous United States to the prediction of the null hypothesis, we
describe changes in the frequency of the most thermally stressing events,
information that is not contained in the trend of mean temperature.
Averaging over these sites and over 365 calendar days permits extraction of
statistically significant results from noisy data.  The deviation of our
metric from the value predicted by the null hypothesis is a measure of the
change in frequency of extreme events.  It also permits defining
``equivalent warming rates'' that describe trends in daily minima or maxima
distinct from the mean rate of global warming.  Comparison of these
equivalent warming rates to the actual mean global warming rate compares (at
the sites for which we have data) the rate of extreme events to that in the
null hypothesis.
\section{Methods}
\label{sec:methods}
\subsection{Data}
The data in this study come from the United States Historical Climatology
Network \citep{USHCN}, a set of 1218 weather monitoring stations across the
48 contiguous United States, selected for quality by the National Climatic
Data Center.
The records include daily minimum and maximum temperatures, in whole degrees
Fahrenheit, through the year 2014 and, for some sites, going back as far as
1850.  We necessarily retain $^\circ$F ($1^{\,\circ}$C = $1.8^{\,\circ}$F)
throughout the analysis because conversion to whole $^\circ$C would lose
temperature resolution, while conversion to decimal $^\circ$C would produce
nearly meaningless values in the tenths place.

\begin{figure}[h!]
	\includegraphics[width=0.5\textwidth]{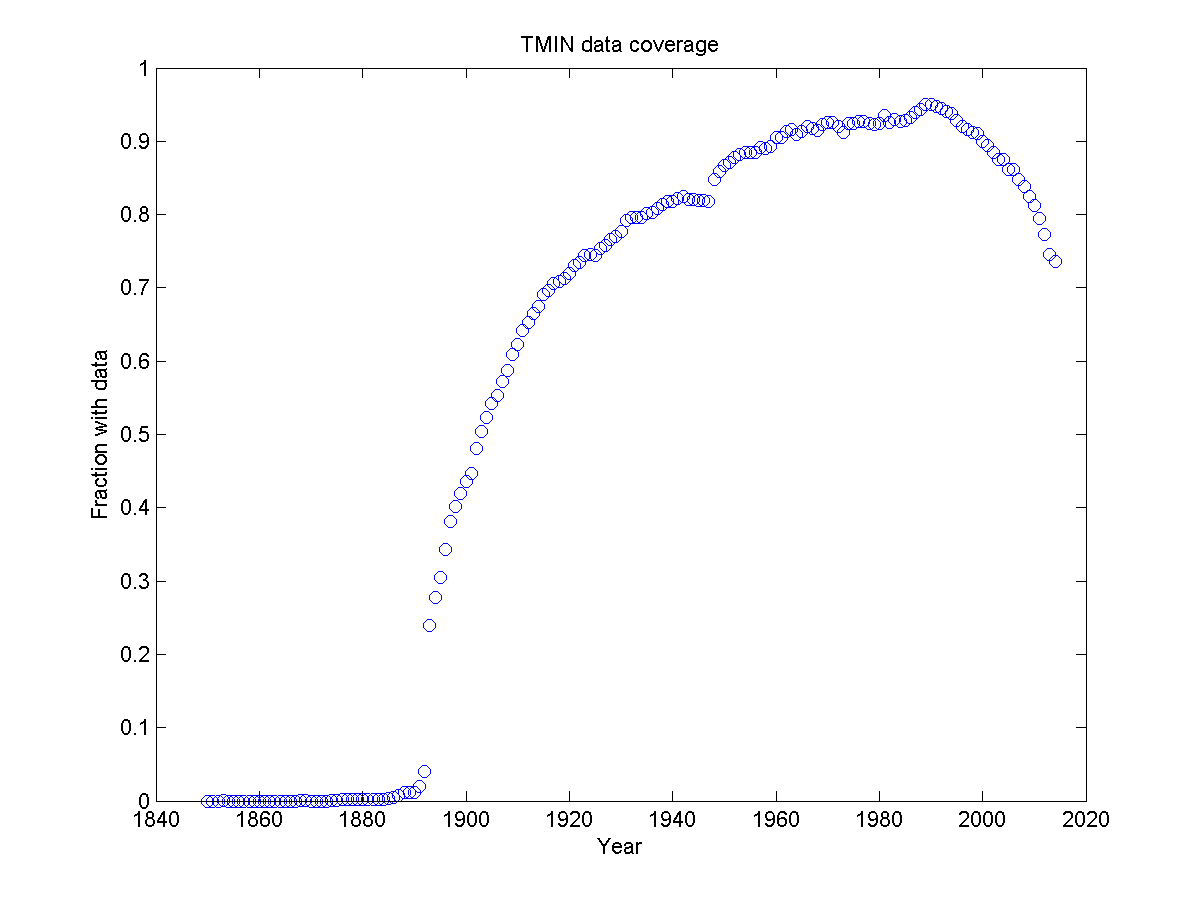}
	\includegraphics[width=0.5\textwidth]{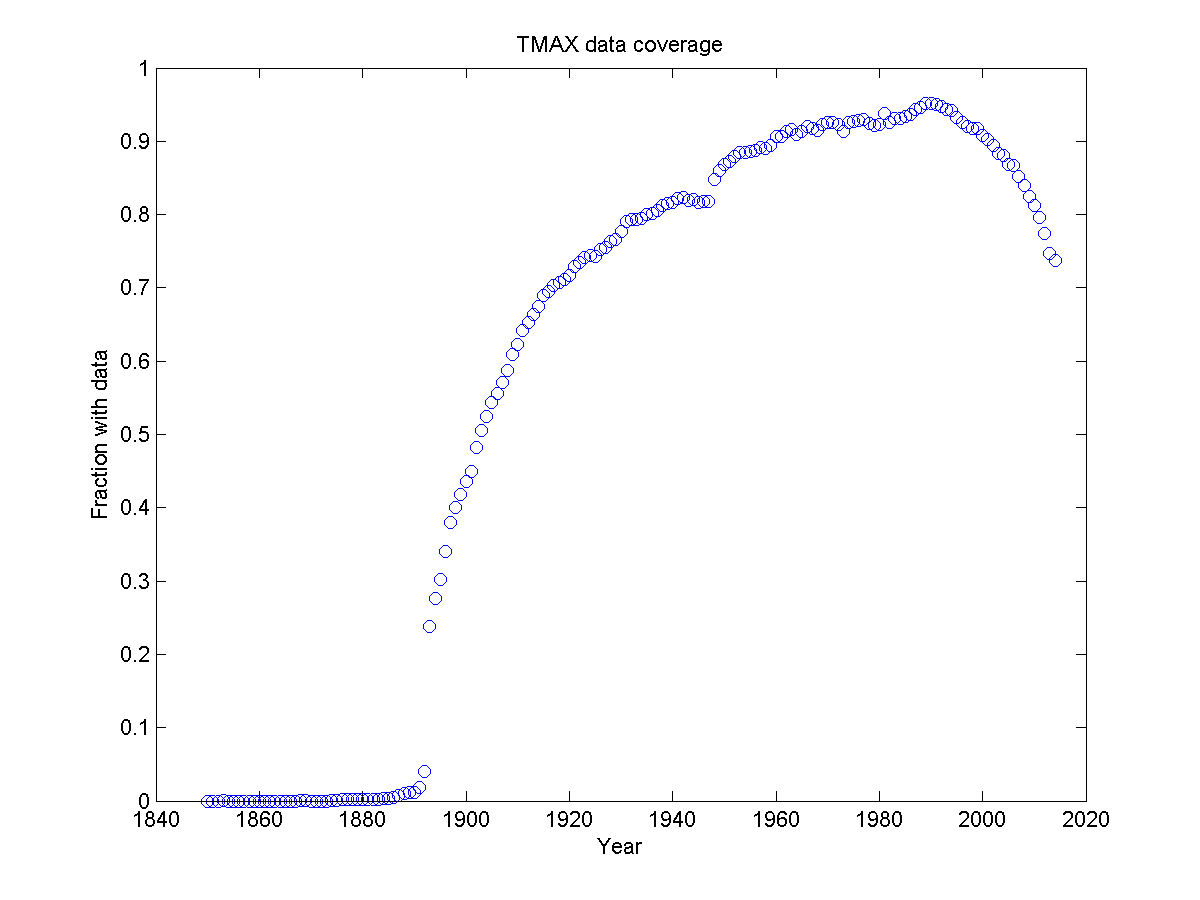}
	\caption{Annual data coverage from 1850 to 2014, as a fraction of 
all $1218\times365$ possible data points each year.}
	\label{fig:1}
\end{figure}

A year of full coverage would have one data point for each of 1218 stations
on each of 365 days (February 29 data are ignored) for a total of 444,570
points.  The actual data coverage is incomplete, especially for the early
years.  Approximately 40.96\% of minima and 40.95\% of maxima are missing.
An additional 0.47\% of minima and 0.43\% of maxima were flagged as suspect 
y the NCDC for failing its consistency and homogeneity tests.  As the
flagged data make up a negligible portion of the dataset, we were able
to use the NCDC's strictest criteria in selecting data for quality, ignoring
data with any flag whatsoever, without compromising data coverage.  Because
of these omissions, the overall data coverage from 1850 to 2014 is about
58\%, with most of the missing data occurring in early years.
Fig.~\ref{fig:1} shows the fraction of these records actually extant each
year.  Annual coverage first exceeds 20\% in 1893 and remains $> 20\%$
thereafter.  We therefore analyzed only data from 1893 onward; the overall
data coverage for 1893--2014 is 79\%.

\subsection{The Metric}
Many metrics have been developed to describe the frequency and severity of
extreme climate \citep{F02,KT03,M06}.  Most of these metrics involve
necessarily arbitrary thresholds and definitions of extreme events, such as
reference to particular percentiles in distributions of weather parameters
or recurrence periods.  We wish a metric that has no parameters or
thresholds with arbitrary quantitative values and that produces a
nonparametric statistical measure of the frequency of extreme events.  This
metric should permit ready comparison to the null hypothesis of a stationary
climate, and its difference from its value in the null hypothesis should
be interpretable as a nominal rate of warming.  This nominal rate describes
the change in frequency of extreme events, and is not the same as the mean
warming rate; the difference in these rates is a measure of change in the
severity of extreme events, distinct from the trend attributable to the mean
warming itself.

The setting of a new maximum or minimum temperature record, for a given
location and a given calendar day, may be defined as an extreme event.
There are no arbitrary parameters in this definition, and the statistic
that compare the frequency of such events to its value for the null
hypothesis of a stationary climate is nonparametric.  The metric is
independent of the underlying temperature distribution and accommodates
nonuniform data coverage.

We define two time series: $\eta(t)$ (the ``high index'') and $\lambda(t)$
(the ``low index''):  Let $s\in\{1,2,\hdots,1218\}$ denote a weather
monitoring station, $d\in\{1,2,\hdots,365\}$ denote a calendar day, and the
whole number $t$ be the number of years since 1850 (the earliest year with a
measurement at any station in our data set).  $T(s,d,t)$ denotes the maximum
temperature measured at site $s$, day $d$ and year $t$. The high index over
these sites is defined as
\begin{equation}
\label{eta}
\eta(t)=\frac{\sum_{s}\sum_{d}K(s,d,t)L(s,d,t)M(s,d,t)}
{\sum_{s}\sum_{d}M(s,d,t)}
\end{equation}
where 
\begin{equation}
\label{M}
M(s,d,t)=\begin{cases}1 & \text{There is a measured high temperature at }
(s,d,t) \\ 0 & \text{otherwise,}\end{cases}
\end{equation}
\begin{equation}
\label{L}
L(s,d,t)=\sum_{t'=0}^t M(s,d,t^\prime),
\end{equation}
and
\begin{equation}
\label{K}
K(s,d,t)=\begin{cases}1 & T(s,d,t)>T(s,d,t^\prime)\ \text{for all } t^\prime<t \\
\frac{1}{n} & T(s,d,t)\text{ is an $n$-way tie for maximum among }t^\prime\leq t
\\ 0 & \text{otherwise.} \end{cases}
\end{equation}
The ``low index" $\lambda(t)$ is defined analogously, replacing $<$ with $>$
in the temperature test of Eq.~\ref{K} and letting $T$ denote the minimum
temperature rather than the maximum.  Even when there are gaps, our metric
utilizes the information contained in all extant data because the likehood,
in the null hypothesis of no climate change, that a new measurement will set
a record is still $1/L(s,d,t)$, however many prior data may be missing.

These metrics measure the frequency of extreme (defined as
record-setting) temperatures without making any assumptions about the
distribution functions of daily maximum and minimum temperatures.  In the
null hypothesis of no warming, the maximum (or minimum) temperature
measurement occurs at year $t$ with probability $1/L(s,d,t)$, which becomes
the expected value of $K(s,d,t)$, so in Eq.~\ref{eta} the numerator's
summand reduces, on average, to $M(s,d,t)$ and $\eta(t)$ and $\lambda(t)$
are, on average, unity, as desired.  In an evolving climate the metrics'
differences from unity are useful nondimensional parametrizations of climate
change. 
As the first measurement for any series is a new record by definition,
$\lambda(1)\equiv\eta(1)\equiv1$, independent of the temperature.  In the
following analysis of these time series, this initial unity value that
carries no information is ignored to maximize sensitivity to climate trends.

$K$ is defined to account for ties for the record, which may occur because
the data are discrete. If year $t$ is
$n$-way tied for the record, there is a probability $1/n$ that the record
(for the period 0--$t$) was indeed year $t$ rather than another of the tied
years.  In the null hypothesis of a stationary climate, any year (for which
there are data) is equally likely to hold the record, so $K(s,d,t)$ has
mean value $1/L(s,d,t)$. On average, then, $\eta(t)$ and $\lambda(t)$ would
equal unity for all years $t$.  Higher values indicate more frequent
record-setting, while lower values indicate less frequent record-setting. 

The indices can be computed for arbitrary subsets of the 1218 sites and
arbitrary subsets of the calendar year, including single sites and single
calendar days.  Such calculations may reveal season- and region-specific
trends that are otherwise obscured in the nationally averaged indicator.
\subsection{Equivalent Warming Rates}
If we know, or are willing to make some assumptions about, the distribution
of daily minima and maxima about their (dependent on site and calendar day)
means, we can describe the difference in the measured indices $\eta(t)$ and
$\lambda(t)$ from unity (their values for a stationary climate) as an
equivalent warming rate.  To do this, we perform a Monte Carlo simulation
of a uniformly warming climate for the period 1850--2014, using the observed
mean warming rate of $0.010^{\,\circ}\text{F/y} = 0.0055^{\,\circ}$C/y
\citep{H06,HRSL10,IPCC5AR}, assuming Gaussian distributions of daily minimum
and maximum temperatures.  The average (over sites and calendar days)
standard deviations of the distributions of daily lows and highs are
$8.15^{\,\circ}$F and $8.96^{\,\circ}$F respectively.  The results are shown
in Table~\ref{tab:ewr} for several values of assumed standard deviations.
To a good approximation, $1 - \overline{\lambda}$ and $\overline{\eta} - 1$
are inversely proportional to the assumed width of the distribution because
they are the first terms in expansions in powers of the ratio of cumulative
temperature change to the width.


\begin{table}
\centering
\caption{Indices of Lows and Highs in Simulated Warming of $0.010^{\,\circ}$F/y}
\begin{tabular}{|c|c|c|}
\hline
Standard Deviation & $\overline{\lambda}$ & $\overline{\eta}$ \\
\hline
$5^{\,\circ}$F & $0.862 \pm 0.008$ & $1.153 \pm 0.009$ \\
$8.15^{\,\circ}$F & $0.913 \pm 0.005$ & $1.093 \pm 0.005$ \\
$8.96^{\,\circ}$F & $0.921 \pm 0.005$ & $1.085 \pm 0.005$\\
$10^{\,\circ}$F & $0.930 \pm 0.004$ & $1.075 \pm 0.004$ \\
$15^{\,\circ}$F & $0.953 \pm 0.003$ & $1.048 \pm 0.003$ \\
\hline
\end{tabular}
\label{tab:ewr}
\end{table}


We define ``equivalent warming rates'' for a measured mean index of highs
$\overline{\eta}$
\begin{equation}
\left.{dT \over dt}\right\vert_{equiv,highs} \equiv
{\overline{\eta} - 1\over 1.085 - 1} \times 0.0055^{\,\circ}\text{C/y},
\label{ewrhigh}
\end{equation}
and of lows $\overline{\lambda}$
\begin{equation}
\left.{dT \over dt}\right\vert_{equiv,lows} \equiv
{1 - \overline{\lambda} \over 1 - 0.913}
\times 0.0055^{\,\circ}\text{C/y}.
\label{ewrlow}
\end{equation}
These equivalent warming rates are normalized so that they equal the actual
mean warming rate if the indices equal those obtained in the Monte Carlo
simulation.  The ratios of equivalent warming rates to the actual mean
warming rate (the fractions in Eqs.~\ref{ewrhigh}, \ref{ewrlow}) indicate
whether extreme weather excursions, warm or cold, are becoming more or less
frequent than they would be in the idealized model in which the
distributions of temperature remain unchanged except for a constant rate of
upward displacement.
\section{Results}
\label{sec:results}
\begin{figure}[h!]
	\includegraphics[width=0.5\textwidth]{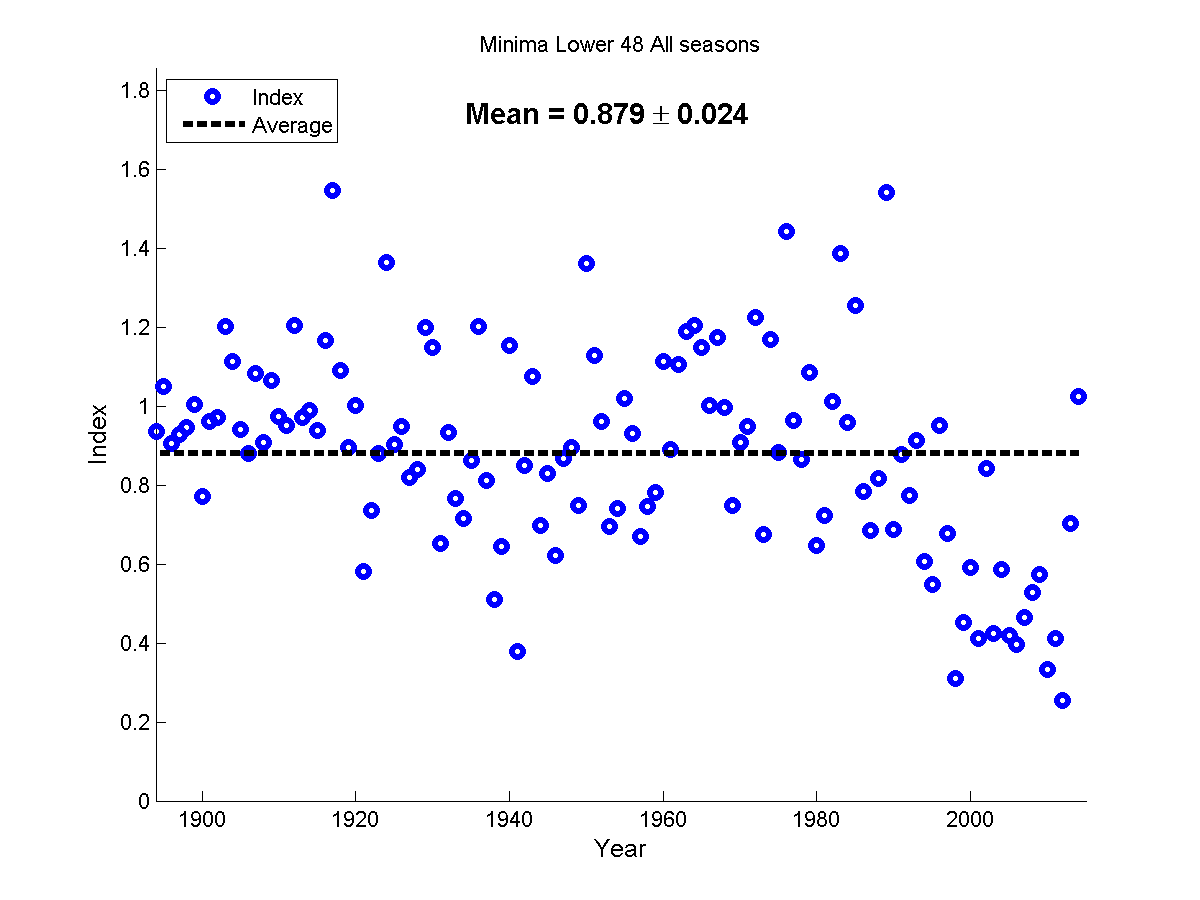}
	\includegraphics[width=0.5\textwidth]{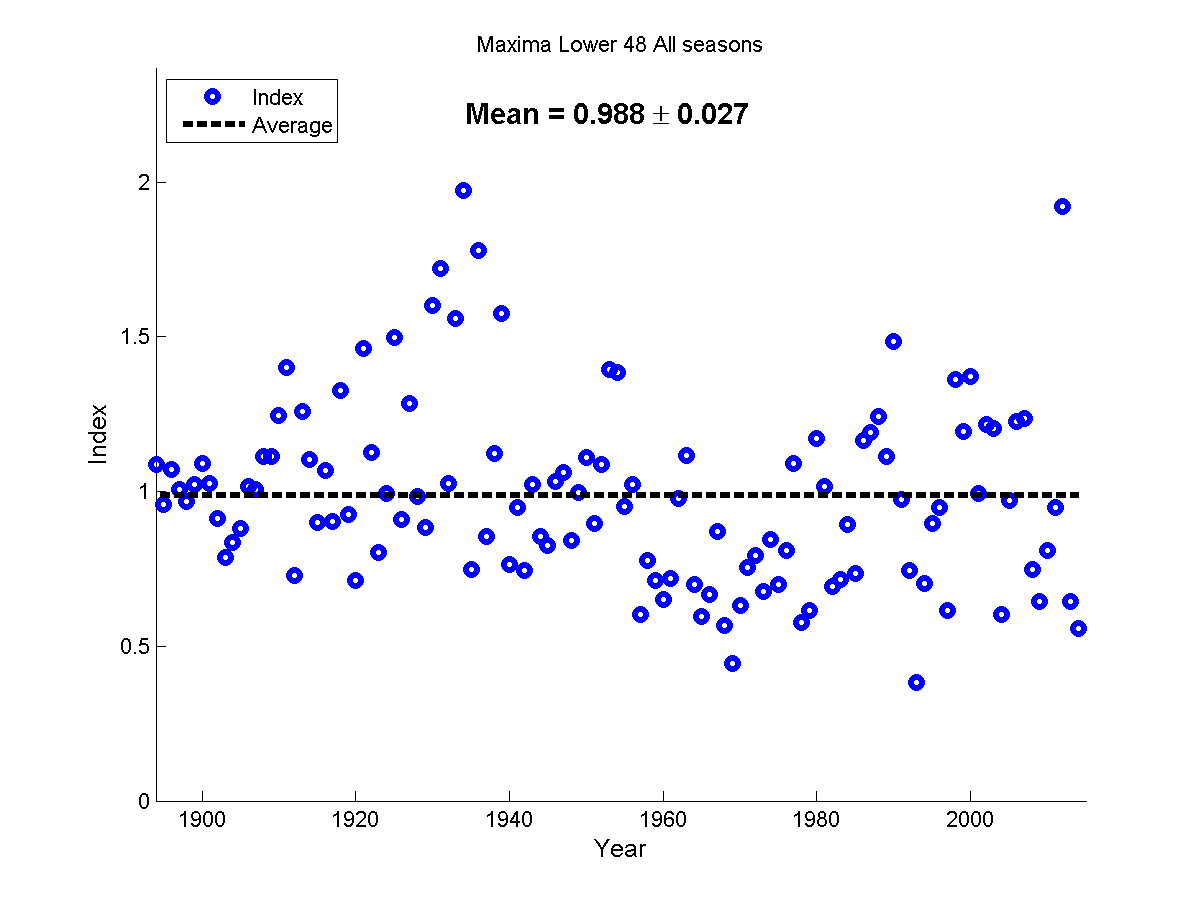}
	\caption{Indices of (a) minimum temperature records $\lambda(t)$ and
(b) maximum temperature records $\eta(t)$ computed over all 1218 sites and
all 365 calendar days, showing means and $\pm 1 \sigma$.}
	\label{fig:2}
\end{figure}

Fig.~\ref{fig:2} shows $\lambda(t)$ and $\eta(t)$ computed
over the 48 contiguous United States and the entire calendar year.
The year-by-year values of $\lambda$ and $\eta$ are scattered, in part
because geographical correlations among the 1218 sites and both day-to-day
(weather system) and year-to-year temporal (ENSO, Decadal {\it etc.\/})
correlations reduce the effectiveness of averaging over the large dataset
in smoothing random fluctuations; the $1218 \times 365$ combinations of site
and calendar day contain many fewer than $1218 \times 365$ independent data.  
However, the means ($\overline{\lambda}$ and $\overline{\eta}$, averaged over
the entire span of data excluding the first year, which is one by definition)
have quite small uncertainties.

A mean index greater (less) than unity occurs if temperature records are set
more (less) often than expected in a stationary climate.  We find that
$\overline{\lambda}<1$ at the $5.0\sigma$ level.
The mean maximum index $\overline{\eta}$ is consistent
with unity (as in the null hypothesis of no climate change) 
All-time temperature minimum records are being set far less frequently than
in a stationary climate, while maximum records are set at the same rate (to
within statistical uncertainty) as in a stationary climate.  This is
consistent with the expectation that global warming tempers cold weather
more than it intensifies hot weather \citep{K93,IPCC5AR}.

These results can be interpreted in terms of the equivalent warming rates
defined in Eqs.~\ref{ewrhigh} and \ref{ewrlow}.  Using the observed standard
deviations
of the daily lows and highs of $8.15^{\,\circ}$F and $8.96^{\,\circ}$F,
respectively, the equivalent warming rates are
\begin{equation}
\left.{dT \over dt}\right\vert_{equiv,highs} = -0.0007 \pm 0.0016^\circ
\text{C/y} \quad \text{and} \quad \left.{dT \over dt}\right\vert_{equiv,lows} =
0.0071 \pm 0.0014\ ^\circ\text{C/y}.
\end{equation}
The frequency of record lows is decreasing at about the same rate as would be
expected from a simple displacement of all temperatures upward in time at
the mean warming rate, but the frequency of record highs is not increasing
at all, and is 
$3.9\sigma$ less than would be predicted by the hypothesis
of uniform upward displacement of all temperatures.

The comparatively rapid warming of the last few decades is also evident in
the $\lambda(t)$ data (Fig.~\ref{fig:2}a).  If a period of rapid warming is
followed by a period of stationary (but elevated, compared to the past)
temperature, all-time record lows will continue to be unusually infrequent
and all-time record highs unusually frequent during that period of
stationary temperature until new records are set.  The period 1990--2014
thus shows an intensification of the depressed $\lambda(t)$ found for the
longer period 1893--2014.  No systematic deviation of $\eta(t)$ from unity
is found for either period.

\begin{figure}[h!]
	\includegraphics[width=0.5\textwidth]{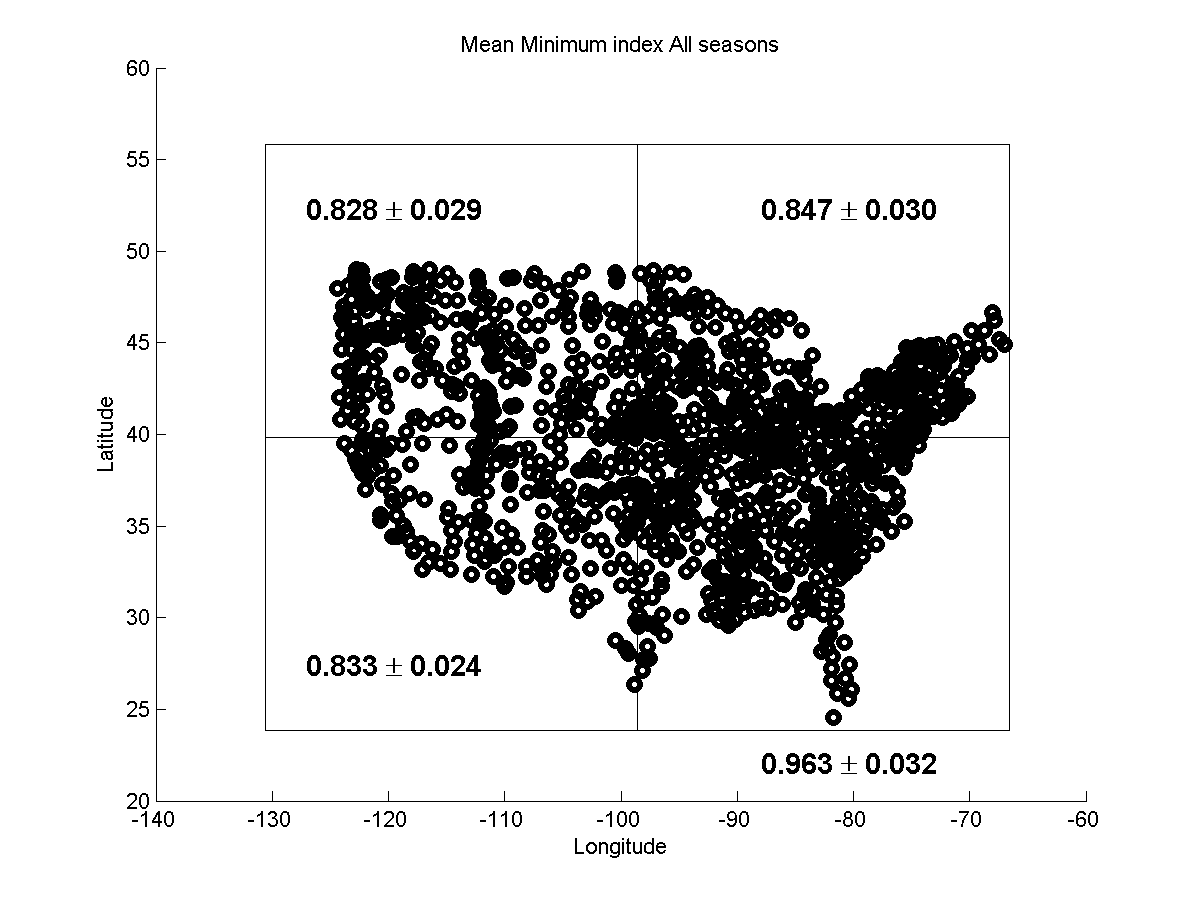}
	\includegraphics[width=0.5\textwidth]{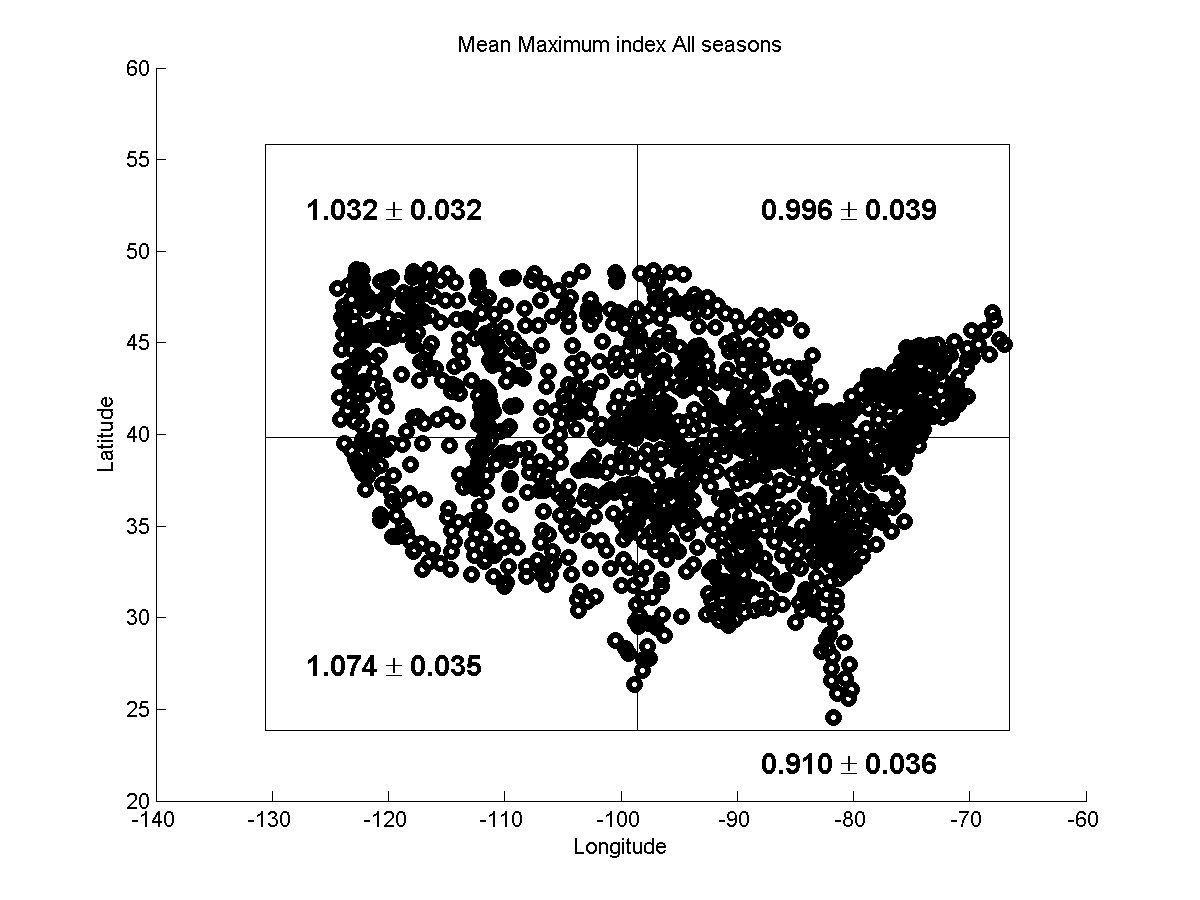}
	\caption{Mean index values and $1\sigma$ uncertainties in individual
quadrants of the contiguous United States, for minima (a) and maxima (b).
Each circle represents a single site.  The sites are broadly distributed,
with some, but not extreme, correlation with population density.}
	\label{fig:3}
\end{figure}

To obtain a more detailed regional picture, we split the 48 contiguous
United States into four quadrants about their geographic center in Lebanon,
Kansas at $39^\circ\ 50^\prime$N, $98^\circ\ 35^\prime$W, and show the
locations of the sites and the results in Fig.~\ref{fig:3}.  The subdivided
values tell a more nuanced story than the nationally averaged values. In
particular, the Southeast shows less warming than the other regions. It is
the only quadrant where $\overline{\lambda}$ falls within $2\sigma$ of
unity; it sets cold records at a rate that is consistent with a stationary
climate, in contrast to the other regions.  In the Southeast
$\overline{\eta}$ is $2.5\sigma$ less than unity; hot records are set at a
rate that is significantly less than that in a stationary climate. This
result is consistent with the Southeastern ``Warming Hole'' identified by
past analyses of climate data and simulations \citep{P04,M12}.
Even though the mean temperature in the Southeast is not decreasing, the
rate of setting record temperature maxima there is suppressed compared to
that suggested by its (nearly zero) mean temperature trend, a difference
between equivalent and actual temperature trends like those found for the
other three quadrants.


We also computed the indices for each season individually, both across the
whole 48 contiguous United States and also subdivided into quadrants.  The
results are shown in
Table~\ref{tab:1} for $\overline{\lambda}$ and Table~\ref{tab:2} for
$\overline{\eta}$.  In these tables ``Winter'' is defined as the first 92
days of the calendar year, and the subsequent seasons as subsequent 91 day
periods.  The Southeastern ``Warming Hole'' is evident in all seasons, with
only a slight (and statistically not significantly different from zero)
decrease in the rate of record minima but a $2.5 \sigma$ ($3.3 \sigma$ in
Spring) decrease in the rate of maxima.  In the other three quadrants the
decreases in minima are generally very significant.  In all quadrants,
seasonal differences are not generally very significant; the trends (or
absence of trends) are found year-round.


\begin{table}
\caption{Mean Minimum Index}
\label{tab:1}
\centering
\begin{tabular}{|l|lllll|}
\hline
Season & Southwest & Northwest & Southeast & Northeast & Lower 48\\
\hline
Winter & $0.820\pm0.049$ & $0.779\pm0.052$ & $0.925\pm0.059$ & $0.842\pm0.056$ & $0.851\pm0.043$\\
Spring & $0.822\pm0.035$ & $0.838\pm0.033$ & $0.987\pm0.043$ & $0.855\pm0.041$ & $0.890\pm0.030$\\
Summer & $0.804\pm0.032$ & $0.773\pm0.030$ & $0.970\pm0.050$ & $0.811\pm0.037$ & $0.853\pm0.031$\\
Fall & $0.886\pm0.040$ & $0.925\pm0.064$ & $0.970\pm0.063$ & $0.880\pm0.058$ & $0.921\pm0.045$\\
Whole Year & $0.833\pm0.024$ & $0.828\pm0.029$ & $0.963\pm0.032$ & $0.847\pm0.030$ & $0.879\pm0.024$\\
\hline
\end{tabular}
\end{table}

\begin{table}
\caption{Mean Maximum Index}
\label{tab:2}
\centering
\begin{tabular}{|l|lllll|}
\hline
Season & Southwest & Northwest & Southeast & Northeast & Lower 48\\
\hline
Winter & $1.048\pm0.063$ & $1.093\pm0.061$ & $0.941\pm0.054$ & $1.173\pm0.085$ & $1.058\pm0.047$\\
Spring & $1.142\pm0.066$ & $1.000\pm0.058$ & $0.833\pm0.050$ & $0.941\pm0.056$ & $0.951\pm0.042$\\
Summer & $1.085\pm0.046$ & $1.068\pm0.042$ & $0.886\pm0.074$ & $0.839\pm0.067$ & $0.945\pm0.046$\\
Fall & $1.023\pm0.062$ & $0.966\pm0.056$ & $0.980\pm0.054$ & $1.026\pm0.063$ & $0.997\pm0.042$\\
Whole Year & $1.074\pm0.035$ & $1.032\pm0.032$ & $0.910\pm0.036$ & $0.996\pm0.039$ & $0.988\pm0.027$\\
\hline
\end{tabular}
\end{table}

\section{Discussion}
We have defined a novel, robust and nonparametric measure of changes in the
frequency of record high or low temperatures.  This metric depends only on
records of daily highs and lows.  It differs from previously defined
measures \citep{F02,KT03,M06,SP06,P08} of temperature extremes by an absence
of arbitrarily defined parameters and thresholds.  Offsetting this
advantage, it does not measure the magnitude of extremes, not
differentiating a slight excursion beyond a previous all-time record from a
large excursion.

Our work is most directly comparable to that of \cite{M09}.  They calculated
the numbers of record daily highs and lows as functions of time for a
world-wide, but very unevenly distributed (concentrated in the same region,
the 48 contiguous United States, that we study), database covering the
shorter period 1950--2006.  They found fewer record lows than would be
expected for a stationary climate, but about the same number of record
highs, consistent with our results.  They only considered the total numbers
of temperature records, summed over sites and calendar days, in contrast to
our treatment of data from each site and each day separately.  This
prevented them from extending their time base to earlier years for which
only a fraction of sites have data or to sites with extensive missing data.
Nor does their method permit quantitative averaging over the entire data
period, although this can be approximated by ``eyeball'' from their figures.
Such averaging smooths the scattered data, produces mean metrics with small
statistical uncertainties, and permits quantitative evaluation of these
uncertainties and of the significance of the results.

\section{Conclusions}
Applying our metrics to an extensive (1218 sites) database of daily
temperature records from the 48 contiguous United States over the period
1893--2014, we find that the frequency of all-time lows decreased at a rate
consistent with a uniform increase of temperatures at the rate of mean
global warming.  However, we find no significant change in the frequency of
all-time highs, a result that differs from a modeled result for uniform
warming at the rate of mean global warming by $4.9\sigma$.  We also 
characterize changes in rates of temperature records in the well-known
``Warming Hole'' in the Southeastern United States, finding the same
comparative suppression of record highs as found in the warming quadrants.

The result that global warming increases temperature minima more than
maxima is expected \citep{H06,HRSL10,IPCC5AR}, and is a natural consequence
of the competing processes of energy transport.  In cold weather energy is
transported by radiation, and is subject to blanketing by greenhouse gases.
In hot weather Solar radiation heats the Earth's surface, producing
instability and convective heat transfer that is little affected by
atmospheric opacity.  Although this qualitative effect is well-established,
the fact that there would be no increase in temperature maxima, as measured
by our metric $\eta$, was unexpected.

\clearpage
\bibliography{finkelextremes}   

\begin{thebibliography}{}
\expandafter\ifx\csname natexlab\endcsname\relax\def\natexlab#1{#1}\fi

\bibitem[{Alexander {et~al.}(2006)Alexander, Zhang, Peterson, Caesar, Gleason,
  Tank, Haylock, Collins, Trewin, Rahimzadeh, Tagipour, Kumar, Revadekar,
  Griffiths, Vincent, Stephenson, Burn, Aguilar, Brunet, Taylor, New, Zhai,
  Rustucucci, \& Vazquez-Aguirre}]{A06}
Alexander, L.~V., Zhang, X., Peterson, T.~C., {et~al.} 2006, Global observed
  changes in daily climate extremes of temperature and precipitation, J.
  Geophys. Res., 111, D05109, doi:10.1029/2005JD006290

\bibitem[{Anderson \& Kostinski(2010)}]{AK10}
Anderson, A., \& Kostinski, A. 2010, Reversible record breaking and
  variability: temperature distributions across the globe, J. Appl. Meteor.
  Clim., 49, 1681

\bibitem[{CDIAC(2015)}]{USHCN}
CDIAC. 2015, \url{cdiac.ornl.gov/epubs/ndp/ushcn/background.html}, accessed
  10/10/2015; site will transition September, 2017

\bibitem[{Easterling {et~al.}(1997)Easterling, Horton, Jones, Peterson, Karl,
  Parker, Salinger, Razuvayev, Plummer, Jamason, \& Folland}]{E97}
Easterling, D.~R., Horton, B., Jones, P.~D., {et~al.} 1997, Maximum and minimum
  temperature trends for the globe, Science, 277, 364,
  doi:10.1126/science.277.5324.364

\bibitem[{Frich {et~al.}(2002)Frich, Alexander, Della-Marta, Gleason, Haylock,
  g.~Klein~Tank, \& Peterson}]{F02}
Frich, P., Alexander, L.~V., Della-Marta, P., {et~al.} 2002, Observed coherent
  changes in climatic extremes during the second half of the twentieth century,
  Clim. Res., 19, 193

\bibitem[{Hansen {et~al.}(2010)Hansen, Ruedy, Sato, \& Lo}]{HRSL10}
Hansen, J., Ruedy, R., Sato, M., \& Lo, K. 2010, Global surface temperature
  change, Rev. Geophys., 48, RG4004, doi:10.1029/2010RG000345

\bibitem[{Hansen(2006)}]{H06}
Hansen, J.~{\it et al\/}. 2006, Global temperature change, Proc. Nat. Acad.
  Sci., 103, 14228

\bibitem[{{Intergovernmental Panel on Climate Change}(2013--2014)}]{IPCC5AR}
{Intergovernmental Panel on Climate Change}. 2013--2014, IPCC Fifth Assessment
  Report (IPCC), \url{www.ipcc.ch/report/ar5/index.shtml}

\bibitem[{Karl {et~al.}(1993)Karl, Jones, Knight, Kukla, Plummer, Razuvayev,
  Gallo, Lindsay, Charlson, \& Peterson}]{K93}
Karl, T.~R., Jones, P.~D., Knight, R.~W., {et~al.} 1993, A new perspective of
  the recent global warming: asymmetric trends of daily maximum and minimum
  temperature, Bull. Am. Meteorol. Soc., 74, 1007

\bibitem[{Katz \& Brown(1992)}]{KB92}
Katz, R.~W., \& Brown, B.~G. 1992, Extreme events in a changing climate:
  variability is more important than averages, Clim. Chang., 21, 289,
  doi:10.1007/bf00139728

\bibitem[{Klein~Tank \& K{\"o}nnen(2003)}]{KT03}
Klein~Tank, A. M.~G., \& K{\"o}nnen, G.~P. 2003, Trends in indices of daily
  temperature and precipitation extremes in Europe, 1946--1999, J. Clim., 16,
  3665

\bibitem[{Lewis {et~al.}(2016)Lewis, King, \& Perkins-Kirkpatrick}]{L16}
Lewis, S., King, A., \& Perkins-Kirkpatrick, S. 2016, Defining a new normal for
  extremes in a warming world, Bull. Amer. Meteor. Soc., in press,
  doi:10.1175/BAMS-D-16-0183.1

\bibitem[{Meehl {et~al.}(2012)Meehl, Arblaster, \& Branstator}]{M12}
Meehl, G.~A., Arblaster, J.~M., \& Branstator, G. 2012, Mechanisms contributing
  to the warming hole and the consequent U.~S. East-West differential of heat
  extremes, J. Clim., 25, 6394, doi:10.1175/JCLI-D-11-00655.1

\bibitem[{Meehl {et~al.}(2009)Meehl, Tebaldi, Walton, Easterling, \&
  McDaniel}]{M09}
Meehl, G.~A., Tebaldi, C., Walton, G., Easterling, D., \& McDaniel, L. 2009,
  Relative increase of record high maximum temperatures compared to record low
  minimum temperatures in the U.S., Geophys. Res. Lett., 36, L23701,
  doi:10.1029/2009GL040736

\bibitem[{Moberg {et~al.}(2006)Moberg, Jones, Lister, Walther, Brunet,
  Jacobeit, Alexander, Della-Marta, Luterbacher, Yiou, Chen, Tank, Saladi{\'e},
  Sigr{\'o}, Aguilar, Alexandersson, Almarza, Auer, Barriendos, Begert,
  Bergstr{\"o}m, B{\"o}hm, Butler, Caesar, adn D.~Founda, Gerstengarbe, Micela,
  Maugeri, {\"O}sterlie, Pandzic, Petrakis, Smec, Tolasz, Tuomenvirta, Werner,
  Linderholm, Philipp, Wanner, \& Xopalki}]{M06}
Moberg, A., Jones, P.~D., Lister, D., {et~al.} 2006, Indices for daily
  temperature and precipitation extremes in Europe analyzed for the period
  1901--2000, J. Geophys. Res., 111, D22106, doi:10.1029/2006JD007103

\bibitem[{Nicholls(1995)}]{N95}
Nicholls, N. 1995, Long-term climate monitoring and extreme events, Clim.
  Chang., 31, 231, doi:10.1007/bf01095148

\bibitem[{Pan {et~al.}(2004)Pan, Arritt, Takle, Jr., Anderson, \& Segal}]{P04}
Pan, Z., Arritt, R.~W., Takle, E.~S., {et~al.} 2004, Altered hydrologic
  feedback in a warming climate introduces a warming hole, Geophys. Res. Lett.,
  31, L17109, doi:10.1029/2004/GL020528

\bibitem[{Platova(2008)}]{P08}
Platova, T.~V. 2008, Annual air temperature extrema in the Russian federation
  and their climatic changes, Russ. Meteorol. Hydrol., 33, 735

\bibitem[{Rahmstorf \& Coumou(2011)}]{RC11}
Rahmstorf, S., \& Coumou, D. 2011, Increase of extreme events in a warming
  world, Pub. Natl. Acad. Sci., 108, 17905

\bibitem[{Shmakin \& Popova(2006)}]{SP06}
Shmakin, A.~B., \& Popova, V.~V. 2006, Dynamics of climate extremes in northern
  Eurasia in the late 20th century, Izvestija Atmos. Ocean. Phys., 42, 157

\end{thebibliography}

%
%
\end{document}